\documentclass[a4paper,11pt]{article}

\usepackage[dvips]{graphicx}
\usepackage{amsmath,amssymb}
\usepackage{ulem}
\usepackage{color}

\newcommand{\bXi}{\boldsymbol{\Xi}}
\newcommand{\bX}{\boldsymbol{X}}
\newcommand{\bx}{\boldsymbol{x}}
\newcommand{\bY}{\boldsymbol{Y}}
\newcommand{\by}{\boldsymbol{y}}
\newcommand{\bz}{\boldsymbol{z}}
\newcommand{\br}{\boldsymbol{r}}
\newcommand{\bZ}{\boldsymbol{Z}}
\newcommand{\bM}{\boldsymbol{M}}

\newcommand{\bE}{\boldsymbol{\epsilon}}

\newcommand{\ba}{\boldsymbol{a}}
\newcommand{\bc}{\boldsymbol{c}}
\newcommand{\bv}{\boldsymbol{v}}
\newcommand{\bw}{\boldsymbol{w}}
\newcommand{\bSi}{\Sigma}
\newcommand{\RMMI}{R_{\mathrm{MMI}}}
\newcommand{\SMMI}{S_{\mathrm{MMI}}}
\newcommand{\UMMI}{U_{\mathrm{MMI}}}

\newcommand{\trans}{^{\mathrm{T}}}

\title{Exploration of synergistic and redundant information sharing in static and dynamical Gaussian systems}

\author{Adam B.~Barrett$^{1,2}$\footnote{adam.barrett@sussex.ac.uk}\\\\
\textit{$^{1}$ Sackler Centre for Consciousness Science} and \textit{Department of Informatics}\\
University of Sussex, Brighton BN1 9QJ, UK\\\\
$^{2}$ \textit{Department of Clinical Sciences}, University of Milan, Milan 20157, Italy}

\date{}

\begin{document}

\maketitle

\begin{abstract}
\noindent To fully characterize the information that two `source' variables carry about a third `target' variable, one must decompose the total information into redundant, unique and synergistic components, i.e.~obtain a partial information decomposition (PID). However Shannon's theory of information does not provide formulae to fully determine these quantities. Several recent studies have begun addressing this. Some possible definitions for PID quantities have been proposed, and some analyses have been carried out on systems composed of discrete variables. Here we present the first in-depth analysis of PIDs on Gaussian systems, both static and dynamical. We show that, for a broad class of Gaussian systems, previously proposed PID formulae imply that: (i) redundancy reduces to the minimum information provided by either source variable, and hence is independent of correlation between sources; (ii) synergy is the extra information contributed by the weaker source when the stronger source is known, and can either increase or decrease with correlation between sources. We find that Gaussian systems frequently exhibit net synergy, i.e.~the information carried jointly by both sources is greater than the sum of informations carried by each source individually. Drawing from several explicit examples, we discuss the implications of these findings for measures of information transfer and information-based measures of complexity, both generally and within a neuroscience setting. Importantly, by providing independent formulae for synergy and redundancy applicable to continuous time-series data, we open up a new approach to characterizing and quantifying information sharing amongst complex system variables.
\end{abstract}

\maketitle

\section{Introduction}
\begin{figure*}
\begin{center}
\includegraphics[width=0.8\textwidth]{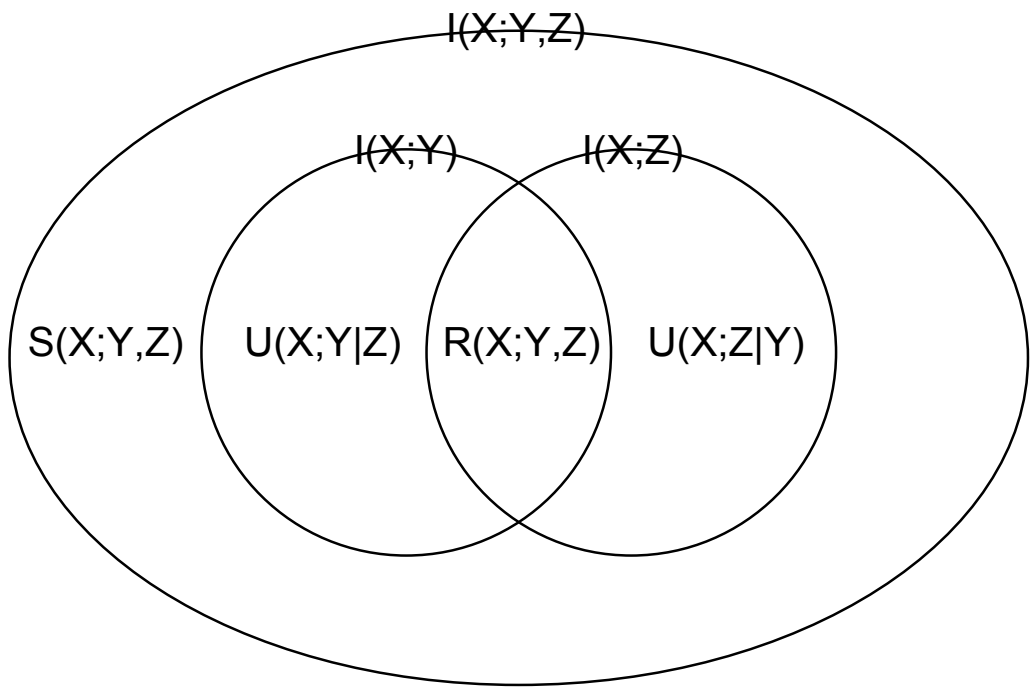}
\end{center}
\caption{The general structure of the information that two `source' variables $\bY$ and $\bZ$ hold about a third `target' variable $\bX$. The ellipses indicate $I(\bX;\bY)$, $I(\bX;\bZ)$ and $I(\bX;\bY,\bZ)$ as labelled, and the four distinct regions enclosed represent the redundancy $R(\bX;\bY,\bZ)$, the synergy $S(\bX;\bY,\bZ)$ and the unique informations $U(\bX;\bY|\bZ)$ and $U(\bX;\bZ|\bY)$ as labelled.} \label{fig:PID}
\end{figure*}

Shannon's information theory \cite{Shannon48} has provided extremely successful methodology for understanding and quantifying information transfer in systems conceptualized as  receiver/transmitter, or stimulus/response \cite{Cover:2006,MacKay:2003}. Formulating information as reduction in uncertainty, the theory quantifies the information $I(\bX;\bY)$ that one variable $\bY$ holds about another variable $\bX$ as the average reduction in the surprise of the outcome of $\bX$ when knowing the outcome of $\bY$ compared to when not knowing the outcome of $\bY$. (Surprise is defined by how unlikely an outcome is, and is given by the negative of the logarithm of the probability of the outcome. This quantity is usually referred to as the mutual information since it is symmetric in $\bX$ and $\bY$.) Recently, information theory has become a popular tool for the analysis of so-called complex systems of many variables, for example, for attempting to understand emergence, self-organisation and phase transitions, and to measure complexity \cite{Prokopenko2009}. Information theory does not however, in its current form, provide a complete description of the informational relationships between variables in a system composed of three or more variables. The information $I(\bX;\bY,\bZ)$ that two `source' variables $\bY$ and $\bZ$ hold about a third `target' variable $\bX$ should decompose into four parts:\footnote{It is our convenient convention of terminology to refer to variables as `sources' and `targets', with $\bY$ and $\bZ$ always being the `sources' that contribute information about the `target' variable $\bX$. These terms relate to the status of the variables in the informational quantities that we compute, and should not be considered as describing the dynamical roles played by the variables.} (i) $U(\bX;\bY|\bZ)$, the unique information that only $\bY$ (out of $\bY$ and $\bZ$) holds about $\bX$; (ii) $U(\bX;\bZ|\bY)$, the unique information that only $\bZ$ holds about $\bX$; (iii) $R(\bX;\bY,\bZ)$, the redundant information that both $\bY$ and $\bZ$ hold about $\bX$; and (iv) $S(\bX;\bY,\bZ)$, the synergistic information about $\bX$ that only arises from knowing both $\bY$ and $\bZ$ (see Figure \ref{fig:PID}). The set of quantities $\{U(\bX;\bY|\bZ),U(\bX;\bZ|\bY),R(\bX;\bY,\bZ),S(\bX;\bY,\bZ)\}$ is called a `partial information decomposition' (PID). Information theory gives us the following set of equations for them:
\begin{eqnarray}
&I(\bX;\bY,\bZ)=U(\bX;\bY|\bZ)+U(\bX;\bZ|\bY)+S(\bX;\bY,\bZ)+R(\bX;\bY,\bZ)\,,& \label{eq:syn1}\\
&I(\bX;\bY)=U(\bX;\bY|\bZ)+R(\bX;\bY,\bZ)\,,& \label{eq:IXYPID}\\
&I(\bX;\bZ)=U(\bX;\bZ|\bY)+R(\bX;\bY,\bZ)\,.& \label{eq:IXZPID}
\end{eqnarray}
However, these equations do not uniquely determine the PID. One can not obtain synergy or redundancy in isolation, but only the `net synergy' or `Whole-Minus-Sum' (WMS) synergy:
\begin{equation}
\mathrm{WMS}(\bX;\bY,\bZ)=:I(\bX;\bY,\bZ)-I(\bX;\bY)-I(\bX;\bZ)=S(\bX;\bY,\bZ)-R(\bX;\bY,\bZ)\,. \label{eq:synminusred}
\end{equation}
An additional ingredient to the theory is required, specifically, a definition that determines one of the four quantities in the PID. A consistent and well-understood approach to PIDs would extend Shannon information theory into a more complete framework for the analysis of information storage and transfer in complex systems.

In addition to the four equations above, the minimal further axioms that a PID of information from two sources should satisfy are: (i) that the four quantities $U(\bX;\bY|\bZ)$, $U(\bX;\bZ|\bY)$, $R(\bX;\bY,\bZ)$ and $S(\bX;\bY,\bZ)$ should always all be greater than or equal to zero; (ii) that redundancy $R(\bX;\bY,\bZ)$ and synergy $S(\bX;\bY,\bZ)$ are symmetric with respect to $\bY$ and $\bZ$ \cite{Williams10}--\cite{Bertschinger13}. Interestingly, several distinct PID definitions have been proposed, each arising from a distinct idea about what exactly should constitute redundancy and/or synergy. These previous studies of PIDs have focused on systems composed of discrete variables. Here, by considering PIDs on Gaussian systems, we provide the first study of PIDs that focuses on continuous random variables.

One might naively expect that for sources and target being jointly Gaussian, the linear relationship between the variables would imply zero synergy, and hence a trivial PID with the standard information theory equations~\eqref{eq:syn1}--\eqref{eq:IXZPID} determining the redundant and unique information. However, this is not the case; net synergy \eqref{eq:synminusred}, and hence synergy, can be positive \cite{Kontoyiannis05,Angelini10}. We begin this study (Section~\ref{sec:prev}) by illustrating the prevalence of jointly Gaussian cases for which net synergy \eqref{eq:synminusred} is positive. Of particular note is the fact that there can be positive net synergy when sources are uncorrelated. After this motivation for the study, in Section~\ref{sec:existing} we introduce three distinct previously proposed PID procedures: (i) that of Williams and Beer \cite{Williams10}; (ii) that of Griffith et al.~\cite{Griffith12,Griffith13} and Bertschinger et al.~\cite{Bertschinger12,Bertschinger13}; and (iii) that of Harder et al.~\cite{Harder12}. In addition to satisfying the minimal axioms above, these PIDs have the further commonality that redundant and unique information depend only on the pair of marginal distributions of each individual source with the target, i.e.~those of $(\bX,\bY)$ and $(\bX,\bZ)$, while only the synergy depends on the full joint distribution of all three variables $(\bX,\bY,\bZ)$. Bertschinger et al.~\cite{Bertschinger13} have argued for this property by considering unique information from a game-theoretic view point. Our key result, that we then demonstrate, is that for a jointly Gaussian system with a univariate target and sources of arbitrary dimension, any PID with this property reduces to simply taking redundancy as the minimum of the mutual informations $I(\bX;\bY)$ and $I(\bX;\bZ)$, and letting the other quantities follow from \eqref{eq:syn1}--\eqref{eq:IXZPID}. This common PID, which we call the MMI (minimum mutual information) PID (i) always assigns the source providing less information about the target as providing zero unique information; (ii) yields redundancy as being independent of the correlation between sources; and (iii) yields synergy as the extra information contributed by the weaker source when the stronger source is known. In Section \ref{sec:Dynamical} we proceed to explore partial information in several example dynamical Gaussian systems, examining (i) the behaviour of net synergy, which is independent of any assumptions on the particular choice of PID, and (ii) redundancy and synergy according to the MMI PID. We then discuss implications for the transfer entropy measure of information flow (Section \ref{sec:TE}), and measures that quantify the complexity of a system via information flow analysis (Section \ref{sec:complexity}). We conclude with a discussion of the shortcomings and possible extensions to existing approaches to PIDs and the measurement of information in complex systems.

This paper provides new tools for exploring information sharing in complex systems, that go beyond what standard Shannon information theory can provide. By providing a PID for triplets of Gaussian variables, it will enable one to study synergy amongst continuous time-series variables, for the first time independently of redundancy. In the Discussion we consider possible application to the study of information sharing amongst brain variables in neuroscience. More generally, there exists possibility of application to complex systems in any realm, e.g.~climate science, financial systems, computer networks, amongst others.



\section{Notation and preliminaries} \label{sec:prelims}
Let $\bX$ be a continuous random variable of dimension $m$. We denote the probability density function by $P_{\bX}(\bx)$, the mean by $\bar{\bx}$, and the $m\times m$ matrix of covariances $\mathrm{cov}(X^i,X^j)$ by $\Sigma(\bX)$. Let $\bY$ be a second random variable of dimension $n$. We denote the $m \times n$ matrix of cross-covariances
$\mathrm{cov}(X^i,Y^j)$ by $\bSi(\bX,\bY)$. We define the `partial covariance' of $\bX$ with respect to $\bY$ as
\begin{equation}
    \bSi(\bX|\bY) =: \bSi(\bX) - \bSi(\bX,\bY) \bSi(\bY)^{-1} \bSi(\bY,\bX)\,. \label{eq:ccxy}
\end{equation}
If $\bX\oplus\bY$ is multivariate Gaussian (we use the symbol `$\oplus$' to denote vertical concatenation of vectors), then the partial
covariance $\bSi(\bX|\bY)$ is precisely the covariance matrix of the
conditional variable $\bX|\bY=\by$, for any $\by$:
\begin{equation} \label{eq:condgauss}
\bX|(\bY=\by) \sim \mathcal{N}[\boldsymbol{\mu}_{\bX|\bY=\by},\bSi(\bX|\bY)]\,,
\end{equation}
where $\boldsymbol{\mu}_{\bX|\bY=\by}=\bar{\bx}+\bSi(\bX,\bY)\bSi(\bY)^{-1}(\by-\bar{\by})$.

Entropy $H$ characterizes uncertainty, and is defined as
\begin{equation} \label{entropy}
H(\bX)=:-\int P_{\bX}(\bx) \log P_{\bX}(\bx)\mathrm{d}^m\bx\,.
\end{equation}
(Note, strictly, Eq.~\eqref{entropy} is the differential entropy, since entropy
itself is infinite for continuous variables. However, considering
continuous variables as continuous limits of discrete variable
approximations, entropy differences and hence information remain
well-defined in the continuous limit and may be consistently
measured using Eq.~\eqref{entropy} \cite{Cover:2006}. Moreover, this equation assumes that $\bX$ has a density with respect to the Lebesgue measure $\mathrm{d}^m\bx$; this assumption is upheld whenever we discuss continuous random variables.) The conditional entropy $H(\bX|\bY)$ is the expected entropy of $\bX$ given $\bY$, i.e.,
\begin{equation}
H(\bX|\bY)=: \int H(\bX|\bY=\by)P_{\bY}(\by)\mathrm{d}^n\by \,.
\end{equation}
The mutual information $I(\bX;\bY)$ between $\bX$ and $\bY$ is the average information, or reduction in uncertainty (entropy), about $\bX$, knowing the outcome of $\bY$:
\begin{equation} \label{eq:mutinf}
I(\bX;\bY)=H(\bX)-H(\bX|\bY)\,.
\end{equation}
Mutual information can also be written in the useful form
\begin{equation} \label{eq:infgen}
I(\bX;\bY)=H(\bX)+H(\bY)-H(\bX,\bY)\,,
\end{equation}
from which it follows that mutual information is symmetric in $\bX$ and $\bY$ \cite{Cover:2006}. The joint mutual information that two sources $\bY$ and $\bZ$ share with a target $\bX$ satisfies a chain rule:
\begin{equation}
I(\bX;\bY,\bZ)=I(\bX;\bY|\bZ)+I(\bX;\bZ)\,, \label{eq:chainrule}
\end{equation}
where the conditional mutual information $I(\bX;\bY|\bZ)$ is the expected mutual information between $\bX$ and $\bY$ given $\bZ$. For $\bX$ Gaussian,
\begin{equation}
H(\bX)=\frac{1}{2}\log [ \det \bSi(\bX)  ] + \frac{1}{2} m \log (2\pi e)\,, \label{eq:Gaussentropy}
\end{equation}
and for $\bX\oplus\bY$ Gaussian
\begin{eqnarray}
H(\bX|\bY)&=&\frac{1}{2}\log [ \det\bSi(\bX|\bY)  ] + \frac{1}{2} m \log (2\pi e)\,,\label{eq:Gausscondent}\\
I(\bX;\bY)&=&\frac{1}{2}\log \left[\frac{ \det\bSi(\bX) }{\det\bSi(\bX|\bY) }\right]\,. \label{eq:Gaussinf}
\end{eqnarray}

For $\bX$ a dynamical variable evolving in discrete time, we denote the state at time $t$ by $\bX_t$, and the infinite past with respect to time $t$ by $\bX_t^{-}=:\bX_{t-1}\oplus \bX_{t-2},\ldots$. The $p$ past states with respect to time $t$ are denoted by $\bX_t^{(p)}=:\bX_{t-1}\oplus \bX_{t-2}\oplus \ldots \oplus \bX_{t-p}$.

\section{Synergy is prevalent in Gaussian systems} \label{sec:prev}

\begin{figure*}
\begin{center}
\includegraphics[width=0.9\textwidth]{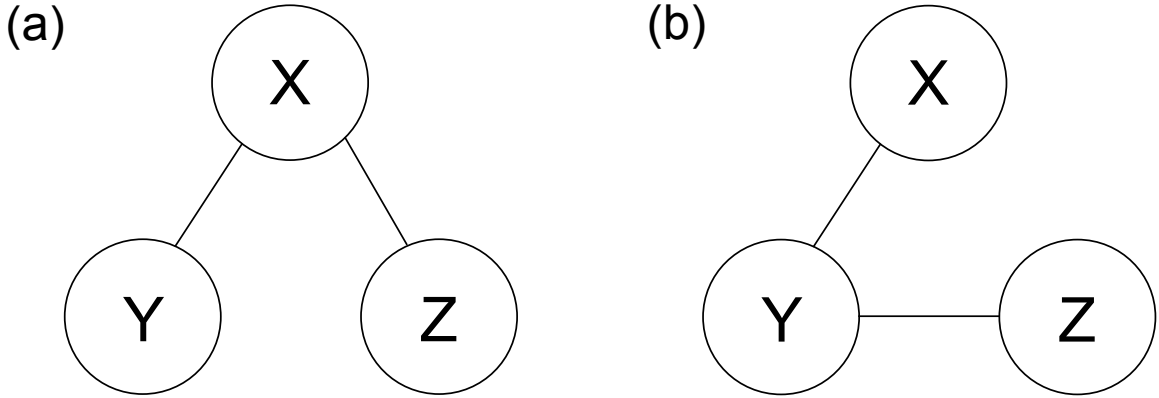}
\end{center}
\caption{The correlational structure of two example systems of univariate Gaussian variables for which $Y$ and $Z$ exhibit positive net synergy with respect to information about $X$. Variables are shown as circles, and the variables that are correlated are joined by lines. (a) $Y$ and $Z$ are uncorrelated and yet show synergy. (b) $X$ and $Z$ are uncorrelated and yet $Z$ contributes synergistic information about $X$ in conjunction with $Y$. See main text for details.} \label{fig:static_eg}
\end{figure*}

In this section we demonstrate the prevalence of synergy in jointly Gaussian systems, and hence that the PIDs for such systems are typically non-trivial. We do this by computing the `Whole-Minus-Sum' (WMS) net synergy, i.e.~synergy minus redundancy \eqref{eq:synminusred}. Since the axioms for a PID impose that $S$ and $R$ are greater than or equal to zero, this quantity provides a lower bound on synergy, and in particular a sufficient condition for non-zero synergy is $\mathrm{WMS}(\bX;\bY,\bZ)>0$. Some special cases have previously been considered in \cite{Kontoyiannis05,Angelini10}. Here we consider, for the first time, the most general three-dimensional jointly Gaussian system $(X,Y,Z)^{\mathrm{T}}$ (here we use normal rather than bold type face for the random variables since they are one-dimensional). Setting means and variances of the individual variables to 0 and 1 respectively preserves all mutual informations between the variables, and so without loss this system can be specified with a covariance matrix of the form
\begin{equation}
\Sigma=\left( \begin{array}{ccc}
1 & a & c \\
a & 1 & b \\
c & b & 1
\end{array} \right)\,, \label{eq:SigmaGauss}
\end{equation}
where $a$, $b$ and $c$ satisfy $|a|,|b|,|c| < 1$, and
\begin{equation}
2abc-a^2-b^2-c^2+1>0
\end{equation}
(a covariance matrix must be non-singular and positive definite).

Using \eqref{eq:ccxy} and \eqref{eq:Gaussinf}, the mutual informations between $X$ and $Y$ and $Z$ are given by
\begin{eqnarray}
I(X;Y)&=&\frac{1}{2}\log \left( \frac{1}{1-a^2} \right)\,, \label{eq:IXYGauss}\\
I(X;Z)&=&\frac{1}{2}\log \left( \frac{1}{1-c^2} \right)\,, \label{eq:IXZGauss}\\
I(X;Y,Z)&=&\frac{1}{2}\log\left( \frac{1-b^2}{1-(a^2+b^2+c^2)+2abc}\right)\,,\label{eq:IXYZGauss}
\end{eqnarray}
and thus the general formula for the net synergy is
\begin{equation}
\mathrm{WMS}(X;Y,Z)=\frac{1}{2}\log \left[ \frac{(1-a^2)(1-b^2)(1-c^2)}{1-(a^2+b^2+c^2)+2abc}\right]\,.
\end{equation}
This quantity is often greater than zero. Two specific examples illustrate the prevalence of net synergy in an interesting way. Consider first the case $a=c$ and $b=0$, i.e. the sources each have the same correlation with the target, but the two sources are uncorrelated [see Figure \ref{fig:static_eg}(a)]. Then there is net synergy since
\begin{equation}
\mathrm{WMS}(X;Y,Z)=\frac{1}{2}\log\left( \frac{1-2a^2+a^4}{1-2a^2} \right)>0\,.
\end{equation}
It is remarkable that there can be net synergy when the two sources are not correlated. However, this can be explained by the concave property of the logarithm function. If one instead quantified information as reduction in covariance, the net synergy would be zero in this case. That is, if we were to define $I_{\bSi}(X;Y)=:\bSi(X) - \bSi(X|Y)$ etc., and $\mathrm{WMS}_{\Sigma}=:I_{\bSi}(X;Y,Z)-I_{\bSi}(X;Y)-I_{\bSi}(X;Z)$, then we would have
\begin{equation}
\mathrm{WMS}_{\Sigma}(X;Y,Z)=\frac{(a^2+c^2)b^2-2abc}{1-b^2},
\end{equation}
which gives the output of zero whenever the correlation $b$ between sources is zero. This is intuitive: the sum of the reductions in covariance of the target given each source individually equals the reduction in covariance of the target given both sources together, for the case of no correlation between sources. There is net synergy in the Shannon information provided by the sources about the target because this quantity is obtained by combining these reductions in covariance non-linearly via the concave logarithm function. This suggests that perhaps $I_{\bSi}$ would actually be a better measure of information for Gaussian variables than Shannon information (although unlike standard mutual information $I_{\bSi}$ is not symmetric). Note that Angelini et al \cite{Angelini10} proposed a version of Granger causality (which is a measure of information flow for variables that are at least approximately Gaussian \cite{Barnett:2009a}) based on straightforward difference of variances without the usual logarithm precisely so that for a linear system the Granger causality from a group of variables equals the sum of Granger causalities from members of the group (see Section\ref{sec:TE} for a recap of the concept of Granger causality).

\begin{figure*}
\begin{center}
\includegraphics[width=0.8\textwidth]{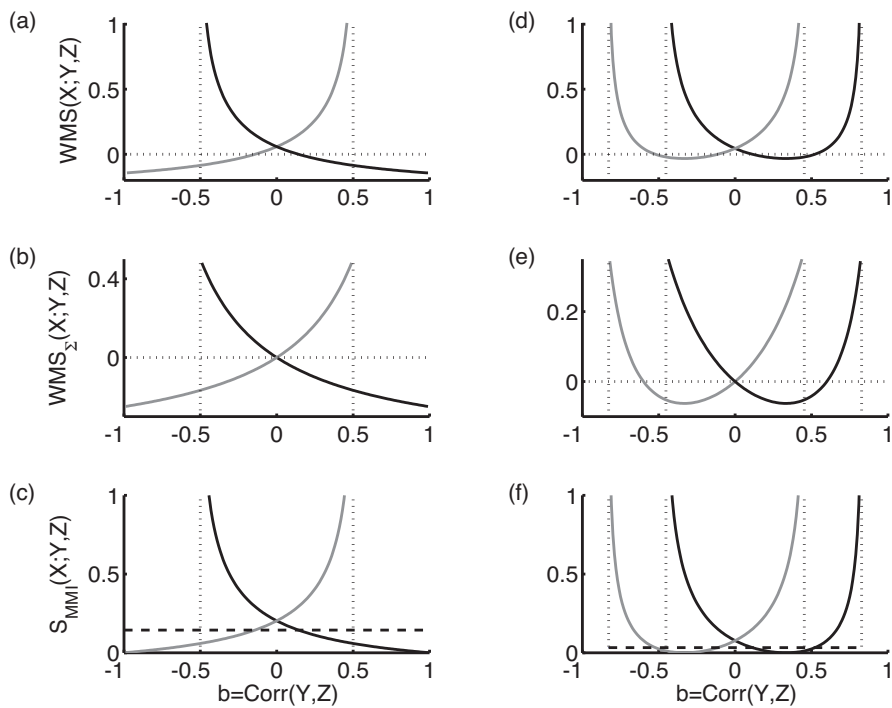}
\end{center}
\caption{Illustrative examples of net synergy WMS and synergy $S_{\mathrm{MMI}}$ between Gaussian variables. (a) Net synergy in Shannon information that sources $Y$ and $Z$ share about the target $X$, as a function of the correlation between $Y$ and $Z$ for (black) correlations between $X$ and $Y$ and $X$ and $Z$ equal and both positive ($a=c=0.5$); (grey) correlations between $X$ and $Y$ and $X$ and $Z$ equal and opposite ($a=-c=0.5$). (b) The same as (a) but using information defined as reduction in variance instead of reduction in Shannon entropy. (c) Synergy according to the MMI PID for the same parameters as (a). Here the dashed line shows redundancy according to the MMI PID, which does not depend on the correlation between $Y$ and $Z$. (d) Example of net synergy as a function of the correlation between $Y$ and $Z$ for (black) correlations between $X$ and $Y$ and $X$ and $Z$ unequal and both positive ($a=0.25$, $c=0.75$); (grey) correlations between $X$ and $Y$ and $X$ and $Z$ unequal and of opposite sign ($a=0.25$, $c=-0.75$). (e) The same as (d) but using information defined as reduction in variance instead of reduction in Shannon entropy. (f) Synergy according to the MMI PID for the same parameters as (d). Here the dashed line shows redundancy according to the MMI PID, which does not depend on the correlation between $Y$ and $Z$. See text for full details of the parameters. In all panels dotted vertical lines indicate boundaries of the allowed parameter space, at which the measures go to infinity, and horizontal dotted lines indicate zero.} \label{fig:wms1}
\end{figure*}

Second, we consider the case $c=0$, i.e. in which there is no correlation between the target $X$ and the second source $Z$ [see Figure \ref{fig:static_eg}(b)]. In this case we have
\begin{equation}
\mathrm{WMS}(X;Y,Z)=\frac{1}{2}\log \left( \frac{1-a^2-b^2+a^2b^2}{1-a^2-b^2} \right)>0\,.
\end{equation}
Hence, the two sources $Y$ and $Z$ exhibit synergistic information about the target $X$ even though $X$ and $Z$ are uncorrelated, and this is modulated by the correlation between the sources $Y$ and $Z$. Although this is perhaps from a naive point of view counter-intuitive, it can be explained by thinking of $Z$ as providing information about why $Y$ has taken the value it has, and from this one can narrow down the range of values for $X$, beyond what was already known about $X$ just from knowing $Y$. Note that in this case there would be net synergy even if one quantified information as reduction in covariance via $I_{\bSi}(X;Y)$ defined above.

Fig.~\ref{fig:wms1}(a,b,d,e) shows more generally how net synergy depends on the correlation between source variables $Y$ and $Z$. For correlations $a$ and $c$ between the two sources and the target being equal and positive, net synergy is a decreasing function of the correlation $b$ between the sources, while for correlations $a$ and $c$ being equal but opposite net synergy is an increasing function of the correlation $b$ between sources [Fig.~\ref{fig:wms1}(a)]. Net synergy asymptotes to infinity as the correlation values approach limits at which the covariance matrix becomes singular. This makes sense because in those limits $X$ becomes completely determined by $Y$ and $Z$. More generally, when $a$ and $c$ are unequal, net synergy is a U-shaped function of correlation between sources [Fig.~\ref{fig:wms1}(d)]. In Fig.~\ref{fig:wms1}(b,e) the alternative measure, $\mathrm{WMS}_{\Sigma}$, of net synergy based on information as reduction in variance is plotted. As described above, this measure behaves more elegantly, always taking the value 0 when the correlation between sources is zero. Taken together these plots show that net redundancy (negative net synergy) does not necessarily indicate a high degree of correlation between source variables.


This exploration of net synergy demonstrates that it would be useful to obtain explicit measures of synergy and redundancy for Gaussian variables. As mentioned in the Introduction, several measures have been proposed for discrete variables \cite{Williams10}--\cite{Bertschinger13}. In the next section we will see that, for a broad class of jointly Gaussian systems, these all reduce essentially to redundancy being the minimum of $I(\bX;\bY)$ and $I(\bX;\bZ)$.

\section{Partial information decomposition on Gaussian systems} \label{sec:existing}
In this section we first revise the definitions of three previously proposed PIDs. We note that all of them have the property that redundant and unique information depend only on the pair of marginal distributions of each individual source with the target, i.e.~those of $(\bX,\bY)$ and $(\bX,\bZ)$, while only the synergy depends on the full joint distribution of all three variables $(\bX,\bY,\bZ)$. Bertschinger et al.~\cite{Bertschinger13} have argued for this property by considering unique information from a game-theoretic view point. We then prove our key result, namely that any PID satisfying this property reduces, for a jointly Gaussian system with a univariate target and sources of arbitrary dimension, to simply taking redundancy as the minimum of the mutual informations $I(\bX;\bY)$ and $I(\bX;\bZ)$, and letting the other quantities follow from \eqref{eq:syn1}--\eqref{eq:IXZPID}. We term this PID the MMI (minimum mutual information) PID, and give full formulae for it for the general fully univariate case considered in Section \ref{sec:prev}. In Section \ref{sec:Dynamical} we go on to apply the MMI PID to dynamical Gaussian systems.

\subsection{Definitions of previously proposed PIDs} \label{sec:prevPIDs}
Williams and Beer's proposed PID uses a definition of redundancy as the minimum information that either source provides about each outcome of the target, averaged over all possible outcomes \cite{Williams10}. This is obtained via a quantity called the specific information. The specific information of outcome $\bX=\bx$ given the random variable $\bY$ is the average reduction in surprise of outcome $\bX=\bx$ given $\bY$:
\begin{equation}
I(\bX=\bx;\bY)=\int \mathrm{d}\by \,p(\by|\bx)\left[ \log \frac{1}{p(\bx)}-\log \frac{1}{p(\bx|\by)} \right]\,.
\end{equation}
The mutual information $I(\bX;\bY)$ is recovered from the specific information by integrating it over all values of $\bx$. Redundancy is then the expected value over all $\bx$ of the minimum specific information that $\bY$ and $\bZ$ provide about the outcome $\bX=\bx$:
\begin{equation}
R(\bX;\bY,\bZ)=\int \mathrm{d}\bx \, p(\bx) \min_{\bXi\in\{ \bY,\bZ\}} I(\bX=\bx;\bXi)\,.
\end{equation}
\\

Griffith et al.~\cite{Griffith12,Griffith13}
consider synergy to arise from information that is not necessarily present given the marginal distributions of source one and target $(\bX,\bY)$ and source two and target $(\bX,\bZ)$. Thus
\begin{equation}
S(\bX;\bY,\bZ)=:I(\bX;\bY,\bZ)-\mathcal{U}(\bX;\bY,\bZ) \label{eq:synunion}
\end{equation}
where
\begin{equation}
\mathcal{U}(\bX;\bY,\bZ)=:\min_{(\tilde{\bX},\tilde{\bY},\tilde{\bZ})}I(\tilde{\bX};\tilde{\bY},\tilde{\bZ})\,, \label{eq:GriffithPID}
\end{equation}
and $\tilde{\bX}$, $\tilde{\bY}$ and $\tilde{\bZ}$ are subject to the constraints $P_{\tilde{\bX},\tilde{\bY}}=P_{\bX,\bY}$ and $P_{\tilde{\bX},\tilde{\bZ}}=P_{\bX,\bZ}$. The quantity $\mathcal{U}(\bX;\bY,\bZ)$ is referred to as the `union information' since it constitutes the whole information minus the synergy. Expressed alternatively, $\mathcal{U}(\bX;\bY,\bZ)$ is the minimum joint information provided about $\bX$ by an alternative $\bY$ and $\bZ$ with the same relations with $\bX$ but different relations to each other.  Bertschinger et al.~\cite{Bertschinger13} independently introduced identically the same PID, but starting from the equation
\begin{equation}
U(\bX;\bY|\bZ)=:\min_{(\tilde{\bX},\tilde{\bY},\tilde{\bZ})}I(\tilde{\bX};\tilde{\bY}|\tilde{\bZ})\,.
\end{equation}
They then derive \eqref{eq:GriffithPID} via the conditional mutual information chain rule  \eqref{eq:chainrule} and the basic PID formulae \eqref{eq:syn1} and \eqref{eq:IXZPID}.
\\

Harder, Salge and Polani's PID \cite{Harder12} define redundancy via the divergence of the conditional probability distribution $P_{\bX|\bZ=\bz}$ for $\bX$ given an outcome for $\bZ$ from linear combinations of conditional probability distributions for $\bX$ given an outcome for $\bY$. Thus, the following quantity is defined:
\begin{equation}
P_{\bX;\bZ=\bz\to \bY}=\mathrm{argmin}_{\by_1,\by_2,\lambda\in[0,1]} \hspace{0.2cm} D_{\mathrm{KL}}\left[ P_{\bX|\bZ=\bz} || \lambda P_{\bX|\bY=\by_1}+(1-\lambda)P_{\bX|\bY=\by_2}\right]\,, \label{eq:pztoy}
\end{equation}
where $D_{\mathrm{KL}}$ is the Kullback-Leibler divergence, defined for continuous probability density functions $P$ and $Q$ by
\begin{equation}
D_{\mathrm{KL}}(P||Q)=:\int P(\bx)\log \left[ \frac{P(\bx)}{Q(\bx)}\right] \mathrm{d}^m\bx\,.
\end{equation}
Then the projected information $I^\pi_{\bX}(\bZ\to \bY)$ is defined as:
\begin{equation}
I^\pi_{\bX}(\bZ\to \bY)=I(\bX;\bZ)-\int \mathrm{d}\boldsymbol{z} p(\bz) D_{\mathrm{KL}}\left[ P_{\bX|\bZ=\bz} || P_{\bX;\bZ=\bz\to \bY}\right]\,, \label{eq:Ixztoy}
\end{equation}
and the redundancy is given by
\begin{equation}
R(\bX;\bY,\bZ)=\min\left\{ I^\pi_{\bX}(\bZ\to \bY), I^\pi_{\bX}(\bY\to \bZ)\right\}\,. \label{eq:HarderR}
\end{equation}
Thus, broadly, the closer the conditional distribution of $\bX$ given $\bY$ is to the conditional distribution of $\bX$ given $\bZ$, the greater the redundancy.


\subsection{The common PID for Gaussians}
While the general definitions of the previously proposed PIDs are quite distinct, one can note that for all of them the redundant and unique informations depend only on the pair of marginal distributions of each individual source with the target, i.e.~those of $(\bX,\bY)$ and $(\bX,\bZ)$. Here we derive our key result, namely the following. Let $X$, $\bY$ and $\bZ$ be jointly multivariate Gaussian, with $X$ univariate and $\bY$ and $\bZ$ of arbitrary dimensions $n$ and $p$. Then there is a unique PID of $I(X;\bY,\bZ)$ such that the redundant and unique informations $R(X;\bY,\bZ)$, $U(X;\bY|\bZ)$ and $U(X;\bZ|\bY)$ depend only on the marginal distributions of $(X,\bY)$ and $(X,\bZ)$. The redundancy according to this PID is given by
\begin{equation}
\RMMI(X;\bY,\bZ)=:\min \left\{ I(X;\bY), I(X;\bZ) \right\}\,. \label{eq:MMI}
\end{equation}
The other quantities follow from \eqref{eq:syn1}--\eqref{eq:IXZPID}, assigning zero unique information to the source providing least information about the target, and synergy as the extra information contributed by the weaker source when the stronger source is known. We term this common PID the MMI (minimum mutual information) PID. It follows that all of the previously proposed PIDs reduce down to the MMI PID for this Gaussian case.
\\

\underline{Proof:} We first show that the PID of Griffith et al.~\cite{Griffith12,Griffith13} (equivalent to that of Bertschinger et al.~\cite{Bertschinger13}) reduces to the MMI PID. Without loss, we can rotate and normalise components of $X$, $\bY$ and $\bZ$ such that the general case is specified by the block covariance matrix
\begin{equation}
\Sigma=\left( \begin{array}{ccc}
1 & \ba^{\mathrm{T}} & \bc^{\mathrm{T}} \\
\ba & I_n & B^{\mathrm{T}} \\
\bc & B & I_p
\end{array} \right)\,,
\end{equation}
where $I_n$ and $I_p$ are respectively the $n$- and $p$-dimensional identity matrices. We can also without loss just consider the case $|\ba|\leq|\bc|$. From \eqref{eq:ccxy} we have
\begin{eqnarray}
\Sigma(X|\bY)&=&1-\ba^{\mathrm{T}}\ba\,, \label{eq:SXY}\\
\Sigma(X|\bZ)&=&1-\bc^{\mathrm{T}}\bc\,, \label{eq:SXZ}
\end{eqnarray}
and hence $I(X;\bY)\leq I(X;\bZ)$. Note then that for a $(\tilde{X},\tilde{\bY},\tilde{\bZ})$
subject to $P_{\tilde{X},\tilde{\bY}}=P_{X,\bY}$ and $P_{\tilde{X},\tilde{\bZ}}=P_{X,\bZ}$
\begin{equation}
I(\tilde{X};\tilde{\bY},\tilde{\bZ})\geq \max \{ I(\tilde{X};\tilde{\bY}),I(\tilde{X};\tilde{\bZ}) \}=\max \left\{ I(X;\bY),I(X;\bZ) \right\}=I(X;\bZ)\,. \label{eq:ineq1}
\end{equation}
Now the covariance matrix of a $(\tilde{X},\tilde{\bY},\tilde{\bZ})$ is given by
\begin{equation}
\tilde{\Sigma}=\left( \begin{array}{ccc}
1 & \ba^{\mathrm{T}} & \bc^{\mathrm{T}} \\
\ba & I_n & \tilde{B}^{\mathrm{T}} \\
\bc & \tilde{B} & I_p
\end{array} \right)\,,
\end{equation}
where $\tilde{B}$ is a $p\times n$ matrix. The residual (partial) covariance of $\tilde{X}$ given $\tilde{\bY}$ and $\tilde{\bZ}$ can thus be calculated using \eqref{eq:ccxy} as
\begin{eqnarray}
\Sigma(\tilde{X}|\tilde{\bY},\tilde{\bZ})&=&1-\left( \ba^{\mathrm{T}} \hspace{0.2cm} \bc^{\mathrm{T}} \right) \left( \begin{array}{cc}
I_n & \tilde{B}^{\mathrm{T}} \\
\tilde{B} & I_p \end{array} \right)^{-1} \left( \begin{array}{c} \ba \\ \bc \end{array} \right) \\
&=& 1 -\bc^{\mathrm{T}}\bc +(\bc^{\mathrm{T}}\tilde{B}-\ba^{\mathrm{T}})(I_n-\tilde{B}^{\mathrm{T}}\tilde{B})^{-1}(\ba-\tilde{B}^{\mathrm{T}}\bc)\,. \label{eq:SXYZ}
\end{eqnarray}
It follows from \eqref{eq:SXYZ} and \eqref{eq:SXZ} that if we could find a $\tilde{B}$ that satisfied $\tilde{B}^{\mathrm{T}}\bc=\ba$, and for which the corresponding $\tilde{\Sigma}$ were a valid covariance matrix, then  $\Sigma(\tilde{X}|\tilde{\bY},\tilde{\bZ})$ would reduce to $\Sigma(X|\bZ)$ and hence we would have $I(\tilde{X};\tilde{\bY},\tilde{\bZ})=I(X;\bZ)$, and thus we would have
\begin{equation}
\mathcal{U}(X;\bY,\bZ)=\max \left\{ I(X;\bY),I(X;\bZ) \right\}\,. \label{eq:unionhyp}
\end{equation}
by \eqref{eq:ineq1} and the definition \eqref{eq:GriffithPID} of $\mathcal{U}$.

We now demonstrate that there does indeed exist a $\tilde{B}$ satisfying $\tilde{B}^{\mathrm{T}}\bc=\ba$ and for which the corresponding $\tilde{\Sigma}$ is positive definite and hence a valid covariance matrix. First note that since $|\ba|\leq|\bc|$ there exists a $\tilde{B}$ satisfying $\tilde{B}^{\mathrm{T}}\bc=\ba$ for which $|\tilde{B}^{\mathrm{T}}\bv|\leq|\bv|$ for all $\bv\in\mathbb{R}^p$. Suppose we have such a $\tilde{B}$. Then the matrix
\begin{equation}
\left( \begin{array}{cc}
I_p & \tilde{B}  \\
\tilde{B}^{\mathrm{T}} & I_n
\end{array} \right)
\end{equation}
is positive definite: For any $\bv\in\mathbb{R}^p$, $\bw\in\mathbb{R}^n$
\begin{eqnarray}
(\bv^{\mathrm{T}} \hspace{0.2cm} \bw^{\mathrm{T}})\left( \begin{array}{cc}
I_p & \tilde{B}  \\
\tilde{B}^{\mathrm{T}} & I_n
\end{array} \right) \left( \begin{array}{c} \bv \\ \bw \end{array} \right)&=&\bv\trans\bv+2\bv\trans\tilde{B}\bw+\bw\trans\bw\\
&\geq& \bv\trans\bv-2\bv\trans\bw+\bw\trans\bw=(\bv-\bw)^2\geq 0\,.
\end{eqnarray}
Since it is also symmetric, it therefore has a Cholesky decomposition:
\begin{equation}
\left( \begin{array}{cc}
I_p & \tilde{B}  \\
\tilde{B}^{\mathrm{T}} & I_n
\end{array} \right) =\left( \begin{array}{cc}
I_p & 0  \\
\tilde{B}^{\mathrm{T}} & P
\end{array} \right)\left( \begin{array}{cc}
I_p & \tilde{B}  \\
0 & P\trans
\end{array} \right)
\end{equation}
where $P$ is lower triangular. Hence, from equating blocks (2,2) on each side of this equation, we deduce that there exists a lower triangular matrix $P$ satisfying
\begin{equation}
\tilde{B}\trans \tilde{B}+PP\trans=I_n\,. \label{eq:chol1}
\end{equation}
We use this to demonstrate that the corresponding $\tilde{\Sigma}$ is positive definite by constructing the Cholesky decomposition for a rotated version of it. Rotating $(X,\bY,\bZ)\to(\bZ,X,\bY)$ leads to the candidate covariance matrix $\tilde{\Sigma}$ becoming
\begin{equation}
\tilde{\Sigma}_{\mathrm{Rot}}=\left( \begin{array}{ccc}
I_p & \bc & \tilde{B} \\
\bc\trans & 1 & \ba^{\mathrm{T}} \\
\tilde{B}\trans & \ba & I_n
\end{array} \right)\,. \label{eq:sigmarot}
\end{equation}
The Cholesky decomposition would then take the form
\begin{equation}
\tilde{\Sigma}_{\mathrm{Rot}}=\left( \begin{array}{ccc}
I_p & \boldsymbol{0} & 0 \\
\bc\trans & q & \boldsymbol{0}^{\mathrm{T}} \\
\tilde{B}\trans & \boldsymbol{r} \trans & S
\end{array} \right)
\left( \begin{array}{ccc}
I_p & \bc & \tilde{B} \\
\boldsymbol{0}\trans & q & \br \\
0 & \boldsymbol{0} & S\trans
\end{array} \right) \label{eq:sigmarotchol}
\end{equation}
where $S$ is a lower triangular matrix, $q$ is a scalar and $\br$ is a vector satisfying
\begin{eqnarray}
\bc\trans\bc+q^2&=&1\,, \label{eq:q}\\
\bc\trans\tilde{B}+q\br&=&\ba\trans\,,\\
\tilde{B}\trans\tilde{B}+\br\trans\br+ SS\trans&=&I_n\,,
\end{eqnarray}
these equations coming respectively from equating blocks (2,2), (2,3) and (3,3) in \eqref{eq:sigmarot} and \eqref{eq:sigmarotchol} (the other block equations are satisfied trivially and don't constrain $S$, $q$ and $\br$). There exists a $q$ to satisfy the first equation since $1-\bc\trans\bc\geq0$ by virtue of it being $\Sigma(X|\bZ)$ \eqref{eq:SXZ} and the original $\Sigma$ being a valid covariance matrix. The second equation is satisfied by $\br=\boldsymbol{0}$ since $\tilde{B}\trans \bc=\ba$. And finally, the third equation is then satisfied by $S=P$, where $P$ is that of \eqref{eq:chol1}. It follows that the Cholesky decomposition exists, and hence $\tilde{\Sigma}$ is a valid covariance matrix, and thus \eqref{eq:unionhyp} holds.

Now, given the definition \eqref{eq:synunion} for the union information and our expression \eqref{eq:unionhyp} for it we have
\begin{equation}
I(X;\bY,\bZ)-S(X;\bY,\bZ)=\max \left\{ I(X;\bY),I(X;\bZ) \right\}\,.
\end{equation}
Thus by the expression \eqref{eq:synminusred} for synergy minus redundancy in terms of mutual information we have
\begin{eqnarray}
R(X;\bY,\bZ)&=&S(X;\bY,\bZ)-I(X;\bY,\bZ)+I(X;\bY)+I(X;\bZ)\\
&=&-\max \left\{ I(X;\bY),I(X;\bZ) \right\}+I(X;\bY)+I(X;\bZ)\\
&=&\min \left\{ I(X;\bY),I(X;\bZ) \right\}\,, \label{eq:MMI2}
\end{eqnarray}
and hence we have reduced this PID to the MMI PID.

Now to show that this is the only PID for this Gaussian case satisfying the given conditions on the marginals of $(X,\bY)$ and $(X,\bZ)$ we invoke Lemma 3 in Ref.~\cite{Bertschinger13}. In Bertschinger et al.'s notation \cite{Bertschinger13}, the specific PID that we have been considering is denoted with tildes, while possible alternatives are written without tildes. It follows from \eqref{eq:MMI2} that the source that shares the smaller amount of mutual information with the target has zero unique information. But according to the Lemma this provides an upper bound on the unique information provided by that source on alternative PIDs. Thus alternative PIDs give the same zero unique information between this source and the target. But according to the Lemma if the unique informations are the same, then the whole PID is the same. Hence, there is no alternative PID. QED.
\\

Note that this common PID does not extend to the case of a multivariate target. For a target with dimension greater than 1, the vectors $\ba$ and $\bc$ above are replaced with matrices $A$ and $C$ with more than one column (these being respectively $\Sigma(\bY,\bX)$ and $\Sigma(\bZ,\bX)$). Then to satisfy \eqref{eq:unionhyp} one would need to find a $\tilde{B}$ satisfying $\tilde{B}\trans C= A$, which does not in general exist. We leave consideration of this more general case to future work.

\subsection{The MMI PID for the univariate jointly Gaussian case}
It is straightforward to write down the MMI PID for the univariate jointly Gaussian case with covariance matrix given by \eqref{eq:SigmaGauss}. Taking without loss of generality $|a|\leq|c|$ we have from \eqref{eq:IXYGauss}--\eqref{eq:IXYZGauss} and \eqref{eq:MMI}:
\begin{eqnarray}
\RMMI(X;Y,Z)&=&I(X;Y)=\frac{1}{2}\log \left( \frac{1}{1-a^2} \right)\,,\label{eq:MMIPID1}\\
\UMMI(X;Y)&=&0\,,\\
\UMMI(X;Z)&=&I(X;Z)-I(X;Y)=\frac{1}{2}\log \left( \frac{1-a^2}{1-c^2}\right)\,,\\
\SMMI(X;Y,Z)&=&\frac{1}{2}\log \left( \frac{(1-b^2)(1-c^2)}{1-(a^2+b^2+c^2)+2abc}\right)\,.\label{eq:MMIPID4}
\end{eqnarray}
It can then be shown that $\SMMI\to \infty$ (and also $\mathrm{WMS}\to\infty$) at the singular limits $b\to ac\pm\sqrt{(1-a^2)(1-c^2)}$, and also that, at $b=a/c$, $\SMMI$ reaches the minimum value of 0. For all in between values there is positive synergy. It is intuitive that synergy should grow largest as one approaches the singular limit, because in that limit $X$ is completely determined by $Y$ and $Z$. On this PID, plots of synergy against correlation between sources take the same shape as plots of net synergy against correlation between sources, because of the independence of redundancy from correlation between sources [Fig.~\ref{fig:wms1}(c,f)]. Thus, for equal (same sign) $a$ and $c$, $\SMMI$ decreases with correlation between sources, for equal magnitude but opposite sign $a$ and $c$, $\SMMI$ increases with correlation between sources, and for unequal magnitude $a$ and $c$ $\SMMI$ has a U-shaped dependence on correlation between sources.

\section{Dynamical systems} \label{sec:Dynamical}
In this section we explore synergy and redundancy in some example dynamical Gaussian systems, specifically multivariate autoregressive (MVAR) processes, i.e., discrete time systems in which the present state is given by a linear combination of past states plus noise.\footnote{These are the standard stationary dynamical Gaussian systems. In fact they are the only stationary dynamical Gaussian systems if one assumes that the present state is a continuous function of the past state \cite{Barrett2010}.}
Having demonstrated (Section \ref{sec:existing}) that the MMI PID is valid for multivariate sources we are able to derive valid expressions for redundancy and synergy in the information that arbitrary length histories of sources contain about the present state of a target. We also compute the more straightforward net synergy.

\begin{figure*}
\begin{center}
\includegraphics[width=0.96\textwidth]{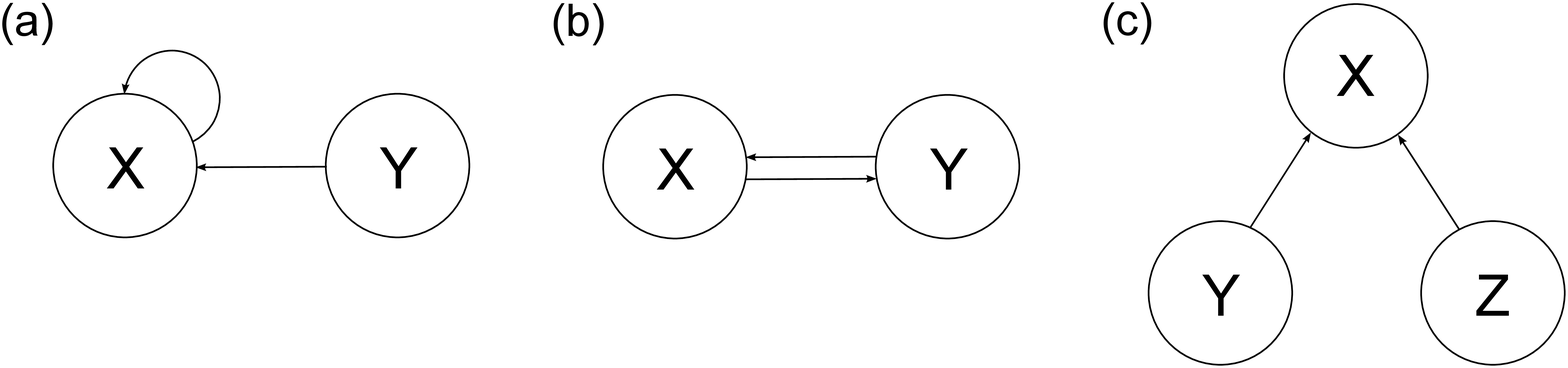}
\end{center}
\caption{Connectivity diagrams for example dynamical systems. Variables are shown as circles, and directed interactions as arrows. The systems are animated as Gaussian MVAR processes of order 1. (a) Example 1. In this system $X$ receives inputs from its own past and from the past of $Y$. There is positive net synergy between the information that the immediate pasts of $X$ and $Y$ provide about the future of $X$, but zero net synergy between the information provided by the infinite pasts of $X$ and $Y$ about the future of $X$. (b) Example 2. In this system there is bidirectional connectivity between $X$ and $Y$. There is zero net synergy between the information provided by the immediate pasts of $X$ and $Y$ about the future of $X$, and negative net synergy (i.e.~positive net redundancy) between the information provided by the infinite pasts of $X$ and $Y$ about the future of $X$. (c) Example 3. Here $Y$ and $Z$ are sources that influence the future of $X$. Depending on the correlation between $Y$ and $Z$, there can be synergy between the information provided by the pasts of $Y$ and $Z$ about the future of $X$ (independent of the length of history considered).} \label{fig:dynamic_eg}
\end{figure*}

\subsection{Example 1: Synergistic two-variable system}
The first example we consider is a two-variable MVAR process consisting of two variables $X$ and $Y$, with $X$ receiving equal inputs from its own past and from the past of $Y$ (see Fig.~\ref{fig:dynamic_eg}(a)). The dynamics are given by the following equations:
\begin{eqnarray}
X_t&=&\alpha X_{t-1}+\alpha Y_{t-1}+\epsilon^X_t\,, \label{eq:X1}\\
Y_t&=&\epsilon^Y_t\,,\label{eq:X2}
\end{eqnarray}
where the $\epsilon$'s are all independent identically distributed Gaussian variables of mean 0 and variance 1. The variables $X$ and $Y$ have a stationary probability distribution as long as $|\alpha|<1$. The information between the immediate pasts of $X$ and $Y$ and the present of $X$ can be computed analytically as follows. First, the stationary covariance matrix $\Sigma(X_t\oplus Y_t)$ satisfies
\begin{equation}
\Sigma(X_t\oplus Y_t)=A\Sigma(X_t\oplus Y_t) A^{\mathrm{T}}+I_{2}\,,
\end{equation}
where $I_{2}$ is the two-dimensional identity matrix and $A$ is the connectivity matrix,
\begin{equation}
A=\left( \begin{array}{cc}
\alpha & \alpha  \\
0 & 0
\end{array} \right)\,.
\end{equation}
This is obtained by taking the covariance matrix of both sides of \eqref{eq:X1} and \eqref{eq:X2}.
Hence
\begin{equation}
\bSi(X_t\oplus Y_t)=\frac{1}{1-\alpha^2}\left( \begin{array}{cc}
1+\alpha^2 & 0  \\
0 & 1-\alpha^2
\end{array} \right)\,.
\end{equation}
The one-lag covariance matrix $\Gamma_1(X_t\oplus Y_t)=:\Sigma(X_t\oplus Y_t,X_{t-1}\oplus Y_{t-1})$ is given by
\begin{equation}
\Gamma_1(X_t\oplus Y_t)=A\bSi(X_t\oplus Y_t)=\frac{\alpha }{1-\alpha ^2}\left( \begin{array}{cc}
1+\alpha ^2 & 1-\alpha ^2  \\
0 & 0
\end{array} \right)\,.
\end{equation}
From these quantities we can obtain the following variances:
\begin{eqnarray}
\bSi(X_t)&=&\frac{1+\alpha ^2}{1-\alpha ^2}\,,\\
\bSi(X_t|X_{t-1})&=&1+\alpha ^2\,,\\
\bSi(X_t|Y_{t-1})&=&\frac{1+\alpha ^4}{1-\alpha ^2}\,,\\
\bSi(X_t|X_{t-1}, Y_{t-1})&=&1\,.
\end{eqnarray}
Then from these we can compute the mutual informations between the present of $X$ and the immediate pasts of $X$ and $Y$:
\begin{eqnarray}
I(X_t;X_{t-1})&=&\frac{1}{2}\log\left( \frac{1}{1-\alpha ^2} \right)\,, \label{eq:Ixx1}\\
I(X_t;Y_{t-1})&=&\frac{1}{2}\log\left( \frac{1+\alpha ^2}{1+\alpha ^4} \right)\,,\\
I(X_t;X_{t-1},Y_{t-1})&=&\frac{1}{2}\log\left( \frac{1+\alpha ^2}{1-\alpha ^2} \right)\,.\label{eq:Ixxy1}
\end{eqnarray}
And thus from these we see that there is net synergy between the immediate pasts of $X$ and $Y$ in information about the present of $X$:
\begin{equation}
\mathrm{WMS}(X_t;X_{t-1},Y_{t-1})=\frac{1}{2}\log\left(1+\alpha ^4 \right)>0\,.
\end{equation}

The infinite pasts of $X$ and $Y$ do not however exhibit net synergistic information about the present of $X$. While $\bSi(X_t|\bX_t^{-})=\bSi(X_t|X_{t-1})$ and $\bSi(X_t|\bX_t^{-},\bY_t^{-})=\bSi(X_t|X_{t-1},Y_{t-1})$, we have $\bSi(X_t|\bY_t^{-})\neq\bSi(X_t|Y_{t-1})$. This is because the restricted regression of $X$ on the past of $Y$ is infinite order:
\begin{equation}
X_t=\sum_{n=1}^\infty \alpha ^nY_{t-n}+\sum_{n=0}^\infty \alpha ^n \epsilon^X_{t-n}\,.
\end{equation}
Hence,
\begin{equation}
\bSi(X_t|\bY_t^{-})=\mathrm{Var}\left(\alpha ^n \epsilon^X_{t-n}\right)=\sum_{n=0}^\infty \alpha ^{2n} = \frac{1}{1-\alpha ^2}\,.
\end{equation}
Therefore
\begin{eqnarray}
I(X_t;\bX_t^{-})&=&\frac{1}{2}\log\left( \frac{1}{1-\alpha ^2} \right)\,,\label{eq:Ixxinf}\\
I(X_t;\bY_t^{-})&=&\frac{1}{2}\log\left( 1+\alpha ^2 \right)\,,\\
I(X_t;\bX_t^{-},\bY_t^{-})&=&\frac{1}{2}\log\left( \frac{1+\alpha ^2}{1-\alpha ^2} \right)\,,\label{eq:Ixxyinf}
\end{eqnarray}
and
\begin{equation}
\mathrm{WMS}(X_t;\bX_t^{-},\bY_t^{-})=0\,.
\end{equation}
Thus the synergy equals the redundancy between the infinite pasts of $X$ and $Y$ in providing information about the present state of $X$.

According to the MMI PID, at infinite lags synergy is the same compared to for one lag, but redundancy is less. We have the following expressions for redundancy and synergy:
\begin{eqnarray}
\RMMI(X_t;X_{t-1},Y_{t-1})&=&\frac{1}{2}\log\left( \frac{1+\alpha ^2}{1+\alpha ^4} \right)\,,\\
\SMMI(X_t;X_{t-1},Y_{t-1})&=&\frac{1}{2}\log\left( 1+\alpha ^2 \right)\,,\\
\RMMI(X_t;\bX_t^{-},\bY_t^{-})=\SMMI(X_t;\bX_t^{-},\bY_t^{-})&=&\frac{1}{2}\log\left( 1+\alpha ^2 \right)\,.
\end{eqnarray}

\subsection{Example 2: An MVAR model with no net synergy}
Not all MVAR models exhibit positive net synergy. The following for example (see Fig.~\ref{fig:dynamic_eg}(b)):
\begin{eqnarray}
X_t&=&\alpha Y_{t-1}+\epsilon^X_t\,,\\
Y_t&=& \beta X_{t-1}+\epsilon^Y_t\,,
\end{eqnarray}
where again the $\epsilon$'s are all independent identically distributed random variables of mean 0 and variance 1, and $|\alpha|,|\beta|<1$ for stationarity. A similar calculation to that for Example 1 shows that the one-lag mutual informations satisfy
\begin{eqnarray}
I(X_t;X_{t-1})&=&0\,,\\
I(X_t;Y_{t-1})&=&\frac{1}{2}\log\left( \frac{1+\alpha^2}{1-\alpha^2\beta^2} \right)\,,\\
I(X_t;X_{t-1},Y_{t-1})&=&\frac{1}{2}\log\left( \frac{1+\alpha^2}{1-\alpha^2\beta^2} \right)\,,
\end{eqnarray}
and thus synergy and redundancy are the same for one-lag mutual information:
\begin{equation}
\mathrm{WMS}(X_t;X_{t-1},Y_{t-1})=0\,.
\end{equation}
For infinite lags one has:
\begin{eqnarray}
I(X_t;\bX_t^{-})&=&\frac{1}{2}\log\left( \frac{1}{1-\alpha^2\beta^2} \right)\,,\\
I(X_t;\bY_t^{-})&=&\frac{1}{2}\log\left( \frac{1+\alpha^2}{1-\alpha^2\beta^2}\right)\,,\\
I(X_t;\bX_t^{-},\bY_t^{-})&=&\frac{1}{2}\log\left( \frac{1+\alpha^2}{1-\alpha^2\beta^2} \right)\,,
\end{eqnarray}
and thus
\begin{equation}
\mathrm{WMS}(X_t;\bX_t^{-},\bY_t^{-})=-\frac{1}{2}\log\left( \frac{1}{1-\alpha^2\beta^2} \right)<0\,,
\end{equation}
so there is greater redundancy than synergy.

For the MMI decomposition we have for 1-lag
\begin{equation}
\RMMI(X_t;X_{t-1},Y_{t-1})=\SMMI(X_t;X_{t-1},Y_{t-1})=0\,,
\end{equation}
while for infinite lags
\begin{eqnarray}
\RMMI(X_t;X_t^{-},Y_t^{-})&=&\frac{1}{2}\log\left( \frac{1}{1-\alpha^2\beta^2} \right)\,,\\
\SMMI(X_t;X_t^{-},Y_t^{-})&=&0\,.
\end{eqnarray}
It is intuitive that for this example there should be zero synergy. All the information contributed by the past of $X$ to the present of $X$ is mediated via the interaction with $Y$, so no extra information about the present of $X$ is gained from knowing the past of $X$ given knowledge of the past of $Y$.

It is interesting to note that for both this example and Example 1 above,
\begin{equation}
\mathrm{WMS}(X_t;X_{t-1},Y_{t-1})>\mathrm{WMS}(X_t;\bX_t^{-},\bY_t^{-})\,. \label{eq:orderdiff}
\end{equation}
That is there is less synergy relative to redundancy when one considers information from the infinite past compared with information from the immediate past of the system. This can be understood as follows. The complete MVAR model is order 1 in each example (that is the current state of the system depends only on the immediate past), so $I(X_t;\bX_t^{-},\bY_t^{-})=I(X_t;X_{t-1},Y_{t-1})$, but restricted effective regressive models of $X$ on just the past of $X$ or just the past of $Y$ are generally of infinite order (that is one can often obtain lower residual noise in $X$ when regressing on the entire infinite past of just $X$ or just $Y$ compared to when regressing on just the immediate past of just $X$ or just $Y$). Hence $I(X_t;\bX_t^{-})\geq I(X_t;X_{t-1})$ and $I(X_t;\bY_t^{-})\geq I(X_t;Y_{t-1})$ for such two variable order 1 MVAR systems. For the two examples, both of these inequalities are strict, and hence the relation \eqref{eq:orderdiff} follows.

An interesting question is whether there exists an MVAR model for two variables $X_t$ and $Y_t$ for which the infinite lag net synergy is greater than zero. It is straightforward to demonstrate that no such system can be found by simple perturbations of the systems considered here. However a full consideration of the most general MVAR model of order greater than 1 is beyond the scope of the present paper. In any case, in the next example, we see that for an MVAR system with three variables, the infinite past of two variables can provide net synergistic information about the future of the third variable.
\\

\subsection{Example 3: Synergy between two variables influencing a third variable}

The third example we consider is an MVAR process with $Y$ and $Z$ being (possibly) correlated sources that are each influencing $X$ (see Fig.~\ref{fig:dynamic_eg}(c)):
\begin{eqnarray}
X_t&=&\frac{1}{\Delta}\left( \alpha Y_{t-1}+\gamma Z_{t-1}+\epsilon^X_t\right)\,,\\
Y_t&=&\epsilon^Y_t\,,\\
Z_t&=&\epsilon^Z_t\,,
\end{eqnarray}
where $\Delta=\sqrt{1+\alpha^2+2\alpha \gamma\rho+\gamma^2}$, and the $\epsilon$'s are Gaussian noise sources all of zero mean, with zero correlation in time, but with instantaneous correlation matrix
\begin{equation}
\Sigma(\boldsymbol{\epsilon})=\left( \begin{array}{ccc}
1 & 0 & 0 \\
0 & 1 & \rho \\
0 & \rho & 1
\end{array} \right)\,.
\end{equation}
Here there is no restriction on connection strengths $\alpha$ or $\gamma$; stationarity is satisfied for all values. Following the same method as in Examples 1 and 2, we have
\begin{equation}
\Sigma(X_t\oplus Y_t \oplus Z_t)=A\Sigma(X_t\oplus Y_t \oplus Z_t) A^{\mathrm{T}}+\Sigma(\boldsymbol{\epsilon})\,,
\end{equation}
and
\begin{eqnarray}
A&=&\frac{1}{\Delta}\left( \begin{array}{ccc}
0 & \alpha & \gamma \\
0 & 0 & 0 \\
0 & 0 & 0
\end{array} \right)\,,\\
\Sigma(X_t\oplus Y_t \oplus Z_t)&=&\left( \begin{array}{ccc}
1 & 0 & 0 \\
0 & 1 & \rho \\
0 & \rho & 1
\end{array} \right)\,,\\
\Gamma_1(X_t\oplus Y_t \oplus Z_t)&=&\frac{1}{\Delta}\left( \begin{array}{ccc}
0 & \alpha+\rho \gamma & \gamma+\rho \alpha \\
0 & 0 & 0 \\
0 & 0 & 0
\end{array} \right)\,.
\end{eqnarray}
From these quantities we can compute the mutual informations:
\begin{eqnarray}
I(X_t;Y_{t-1})&=&\frac{1}{2}\log \left( \frac{1+\alpha ^2+2\alpha \gamma \rho+\gamma ^2}{1+\gamma ^2(1-\rho^2)} \right)\,,\\
I(X_t;Z_{t-1})&=&\frac{1}{2}\log \left( \frac{1+\alpha ^2+2\alpha \gamma \rho+\gamma ^2}{1+\alpha ^2(1-\rho^2)} \right)\,,\\
I(X_t;Y_{t-1},Z_{t-1})&=&\frac{1}{2}\log \left( 1+\alpha ^2+2\alpha \gamma \rho+\gamma ^2 \right)\,. \label{eq:Itotaleg3}
\end{eqnarray}
Hence, assuming without loss of generality that $|\alpha |\leq|\gamma |$,
\begin{eqnarray}
\mathrm{WMS}(X_t;Y_{t-1},Z_{t-1})&=&\frac{1}{2}\log \left( \frac{[1+\alpha ^2(1-\rho^2)][1+\gamma ^2(1-\rho^2)]}{1+\alpha ^2+2\alpha \gamma \rho+\gamma ^2} \right)\,, \label{eq:WMSEg3}\\
\RMMI(X_t;Y_{t-1},Z_{t-1})&=&\frac{1}{2}\log \left( \frac{1+\alpha ^2+2\alpha \gamma \rho+\gamma ^2}{1+\gamma ^2(1-\rho^2)} \right)\,,\\
\SMMI(X_t;Y_{t-1},Z_{t-1})&=&\frac{1}{2}\log \left( 1+\alpha ^2[1-\rho^2] \right)\,. \label{eq:SEg3}
\end{eqnarray}
Note we do not consider the PID for the information provided by the infinite pasts of $Y$ and $Z$ because it is the same as that provided by the immediate pasts for this example.

For the case of no correlation between $Y$ and $Z$, i.e.~$\rho=0$, we have
\begin{equation}
\mathrm{WMS}(X_t;Y_{t-1},Z_{t-1})=\frac{1}{2}\log \left( \frac{[1+\alpha ^2][1+\gamma ^2]}{1+\alpha ^2+\gamma ^2} \right)>0\,,
\end{equation}
i.e.~there is net synergy. For the case $\rho=1$ of $Y$ and $Z$ being perfectly correlated, there is however net redundancy, since
\begin{equation}
\mathrm{WMS}(X_t;Y_{t-1},Z_{t-1})=\frac{1}{2}\log \left( \frac{1}{1+(\alpha +\gamma )^2} \right)<0\,.
\end{equation}
This is a dynamical example in which two uncorrelated sources can contribute net synergistic information to a target. The MMI PID synergy $\SMMI$ behaves in an intuitive way here, increasing with the square of the weaker connection $\alpha$, and decreasing as the correlation $\rho$ between the sources $Y$ and $Z$ increases, and going to zero when $\alpha=0$ or $\rho=1$, reflecting the strength and independence of the weaker link.

Considering this system further for the case $\rho=0$ and $\alpha=\gamma $, for small $\alpha$ the net synergy is approximately $\alpha^4/2$, and for large $\alpha$ the net synergy is approximately $\log (\alpha/\sqrt 2)$ (as stated above $\alpha$ can be arbitrarily large in this model, since the spectral radius i.e. largest absolute value of the eigenvalues of the connectivity matrix, is zero independent of $\alpha$). Hence net synergy can be arbitrarily large. The proportion $\mathrm{WMS}(X_t;Y_{t-1},Z_{t-1})/I(X_t;Y_{t-1},Z_{t-1})$ also grows with connection strength $\alpha$, reaching for example approximately $0.1$ for $\alpha=0.5$.

\section{Transfer entropy} \label{sec:TE}
The net synergy in the example systems of Section \ref{sec:Dynamical} affect transfer entropy and its interpretation. Pairwise (one-lag) transfer entropy is defined as
\begin{equation}
\mathcal{T}^{(1)}_{Y\to X}=: H(X_t|X_{t-1})-H(X_t|X_{t-1},Y_{t-1})\equiv I(X_t;X_{t-1},Y_{t-1})-I(X_t;X_{t-1})\,.
\end{equation}
Typically transfer entropy is interpreted straightforwardly as the information that the past of $Y$ contributes to the present of $X$ over and above that already provided by the past of $X$ \cite{Lizier09}. It has sometimes been implicitly assumed to be less than the lagged mutual information $I(X_t;Y_{t-1})$ for simple linear systems, for example, in constructing measures of the overall causal interactivity of a system \cite{Barrett2010}. However, this is not the case when there is net synergy, since transfer entropy measures the unique information provided by the past of $Y$ plus the synergistic information between the pasts of $X$ and $Y$,
\begin{equation}
\mathcal{T}^{(1)}_{Y\to X}=U(X_t;Y_{t-1}|X_{t-1})+S(X_t;X_{t-1},Y_{t-1})\,,
\end{equation}
whereas the lagged mutual information $I(X_t;Y_{t-1})$ measures the unique information provided by the past of $Y$ plus the redundant information provided by the pasts of $X$ and $Y$:
\begin{equation}
I(X_t;Y_{t-1})=U(X_t;Y_{t-1}|X_{t-1})+R(X_t;X_{t-1},Y_{t-1})\,.
\end{equation}
Specifically for Example 1,
\begin{eqnarray}
\mathcal{T}^{(1)}_{Y\to X}&=&\frac{1}{2}\log\left( 1+\alpha^2 \right)\,,\\
\mathcal{T}^{(1)}_{Y\to X}-I(X_t;Y_{t-1})&=&\frac{1}{2}\log\left( 1+\alpha^4 \right)>0\,.
\end{eqnarray}

The situation can be different when infinite lags are considered:
\begin{equation}
\mathcal{T}^{(\infty)}_{Y\to X}=: H(X_t|\bX_t^{-})-H(X_t|\bX_t^{-},\bY_t^{-})\,.
\end{equation}
For Example 1, considering infinite lags, the transfer entropy $\mathcal{T}^{(\infty)}_{Y\to X}$ and the lagged mutual information $I(X_t;\bY_t^{-})$ are equal because the net synergy between complete past histories of $X$ and $Y$ is zero. From \eqref{eq:Ixx1}--\eqref{eq:Ixxy1} and \eqref{eq:Ixxinf}--\eqref{eq:Ixxyinf} we have
\begin{eqnarray}
\mathcal{T}^{(\infty)}_{Y\to X}&=&\mathcal{T}^{(1)}_{Y\to X}\,,\\
\mathcal{T}^{(\infty)}_{Y\to X}-I(X_t;\bY_t^{-})&=&0\,.
\end{eqnarray}

Conditional transfer entropy $\mathcal{T}^{(\infty)}_{Y\to X|Z}$ (infinite lags) is defined as
\begin{equation}
\mathcal{T}^{(\infty)}_{Y\to X|Z}=: H(X_t|\bX_t^{-},\bZ_t^{-})-H(X_t|\bX_t^{-},\bY_t^{-},\bZ_t^{-})\,.
\end{equation}
It has sometimes been assumed that the conditional transfer entropy is less than non-conditional transfer entropy, i.e. $\mathcal{T}^{(\infty)}_{Y\to X|Z}$ is less than $\mathcal{T}^{(\infty)}_{Y\to X}$ \cite{SethEtal06,Barrett2010}. This is because the pasts of $Y$ and $Z$ might contribute redundant information to the future of $X$, but as for pairwise non-conditional transfer entropy, synergy is usually not considered important for continuous, linear unimodal systems such as those considered in this manuscript. However, for Example 3 this is not always true. Considering the net synergistic case of $\rho=0$, $\alpha=\gamma$,
\begin{eqnarray}
\mathcal{T}_{Y\to X}&=&\frac{1}{2}\log\left( \frac{1+2\alpha^2}{1+\alpha^2} \right)\,,\\
\mathcal{T}_{Y\to X|Z}&=&\frac{1}{2}\log\left( 1+\alpha^2 \right)\,,\\
\mathcal{T}_{Y\to X|Z}-\mathcal{T}_{Y\to X}&=&\frac{1}{2}\log\left( 1+\frac{\alpha^4}{1+2\alpha^2} \right)>0\,.
\end{eqnarray}
Here the number of lags is left unspecified because these quantities are the same for any number of lags. Thus conditional transfer entropy can be affected by synergy even when infinite lags are considered. In this example, because $X$ has no self-connection, and thus the past of $X$ contributes no information to the future of $X$, $\mathcal{T}_{Y\to X}$ reduces to $U(X_t;Y_{t-1}|Z_{t-1})+R(X_t;Y_{t-1},Z_{t-1})$ and $\mathcal{T}_{Y\to X|Z}$ to $U(X_t;Y_{t-1}|Z_{t-1})+S(X_t;Y_{t-1},Z_{t-1})$. Non-conditional minus conditional transfer entropy has been applied to assess the balance between synergy and redundancy (i.e.~net synergy) amongst neuroelectrophysiological variables in \cite{Stramaglia12}.


Since transfer entropy is equivalent to the linear formulation of Granger causality for jointly Gaussian variables \cite{Barnett:2009a}, the above conclusions pertain also to interpretations of Granger causality. Granger causality quantifies the extent to which the past of one variable $Y$ predicts the future of another variable $X$ over and above the extent to which the past of $X$ (and the past of any `conditional' variables) predicts the future of $X$ \cite{Wiener56,Granger69}. In the usual linear formulation, the prediction is implemented using the framework of linear autoregression. Thus, to measure the
Granger causality from `predictor' $Y$ to `predictee' $X$ given conditional variables $\bZ$, one compares the following
multivariate autoregressive (MVAR) models:
\begin{eqnarray}
    X_t & =& A \cdot [\bX^{(p)}_{t}\oplus{\bZ^{(r)}_{t}}] + \bE_t\,, \\
    X_t & = &A' \cdot [\bX^{(p)}_{t}\oplus{\bY^{(q)}_{t}}\oplus{\bZ^{(r)}_{t}}] + \bE'_t\,. \label{eq:reg}
\end{eqnarray}
Thus the `predictee' variable $X$ is regressed firstly on the
previous $p$ lags of itself plus $r$ lags of the conditioning
variables $\bZ$ and secondly, in addition, on $q$ lags of the predictor
variable $Y$ ($p$, $q$ and $r$ can be selected according to the Akaike or
Bayesian information criterion \cite{DingEtal06}). The magnitude of the Granger causality interaction is then given by the logarithm of the ratio
of the residual variances:
\begin{equation}
    \mathcal{F}_{Y\to X |\bZ} =: \log\left( \frac{\bSi(\bE_t)}{\bSi(\bE'_t)}\right) = \log\left( \frac{\bSi( X_t|\bX_t^-,\bZ_t^-)}{\bSi( X_t|\bX_t^-,\bY_t^-,\bZ_t^-)}\right)\,,
    \label{eq:gcu}
\end{equation}
where the final term expresses Granger causality in terms of partial covariances, and hence illustrates the equivalence with transfer entropy for Gaussian variables (up to a factor of 2) \cite{Barnett:2009a}. It follows that pairwise Granger causality $\mathcal{F}_{Y\to X}$ (no conditional variables) should be considered as a measure of the unique (with respect to the past of $X$) predictive power that the past of $Y$ has for the future of $X$ plus the synergistic predictive power that the pasts of $X$ and $Y$ have in tandem for the future of $X$. Meanwhile conditional Granger causality $\mathcal{F}_{Y\to X |\bZ}$ should be considered as a measure of the unique (with respect to the pasts of $X$ and $\bZ$) predictive power that the past of $Y$ has for the future of $X$ plus the synergistic predictive power that the pasts of $X$ and $Y\oplus \bZ$ have in tandem for the future of $X$.

\section{Implications for measures of overall interactivity and complexity} \label{sec:complexity}
The prevalence of synergistic contributions to information sharing has implications for how to sensibly construct measures of overall information transfer sustained in a complex system, or the overall \textit{complexity} of the information transfer.

One such measure is causal density \cite{SethEtal06,Barrett2010,Seth:2011}. Given a set of Granger causality values among elements of a system $\bM$,
a simple version of causal density can be defined as the average of all
pairwise Granger causalities between elements (conditioning on all remaining
elements):
\begin{equation}
\mathrm{cd}(\bM) =: \frac{1}{n(n-1)} \sum_{i \neq j} \mathcal{F}_{M_j\to M_i|\bM_{[ij]}}\,, \label{cd1}
\end{equation}
where $\bM_{[ij]}$ denotes the subsystem of $\bM$ with variables $M_i$ and $M_j$ omitted, and $n$ is the total number of variables.  Causal density provides a principled measure of dynamical complexity inasmuch as elements that are completely independent will score zero, as will elements that are completely integrated in their dynamics. High values will only be achieved when elements behave somewhat differently from each other, in order to contribute novel potential predictive information, and at the same time are globally integrated, so that the potential predictive information is in fact useful \cite{SethEtal06,shanahan:2008}. In the context of the current discussion however, causal density counts synergistic information multiple times, whilst neglecting redundant information. For instance, in Example 3 above, the non-zero contributions to causal density are
\begin{eqnarray}
\mathrm{cd}&=&\frac{1}{6} \left[ \mathcal{F}_{Z\to X|Y} + \mathcal{F}_{Y \to X |Z}\right]\\
&=&\frac{1}{3}\left[ U(X_t;Y_{t-1}|Z_{t-1})+U(X_t;Z_{t-1}|Y_{t-1})+2S(X_t;Y_{t-1},Z_{t-1}) \right]\,.
\end{eqnarray}
In spite of this apparent overcounting of synergistic information, the resultant formula is
\begin{equation}
\mathrm{cd}=\frac{1}{6}\left\{ \log  [1+\alpha^2(1-\rho^2)]  + \log [1+\gamma^2(1-\rho^2)]\right\}\,,
\end{equation}
which is after all a sensible formula for the overall level of transfer of novel predictive information, increasing with connection strengths $\alpha$ and $\gamma$ and decreasing with the correlation $\rho$ between the source variables, and going to zero if either both $\alpha$ and $\gamma$ are zero or if $\rho\to1$.

An alternative to causal density is the global transfer entropy \cite{Lizier10,Barnett13}, $\mathcal{T}_\mathrm{gl}$, defined as
\begin{equation}
\mathcal{T}_\mathrm{gl}(\bM)=:\frac{1}{n} \sum_i \mathcal{T}_{\boldsymbol{M}\to M_i}\,,
\end{equation}
i.e.~the average information flow from the entire system to individual elements. This may be considered a measure of gross past-conditional statistical dependence of the elements of the system, insofar as it vanishes if and only if each system element, conditional on
its own past, does not depend on the past of other system elements. Unlike causal density, this measure assigns equal weight to contributions from unique, redundant and synergistic information flow. However, it is not sensitive to whether the information flow occurs homogeneously or inhomogeneously; it does not care about the distribution amongst sources of the information that flows into the targets. It should thus be interpreted as operationalising a different conceptualisation of complexity to causal density. For Example 3 above, the only non-zero contribution to this global transfer entropy arises from $I(X_t;Y_{t-1},Z_{t-1})$. Thus from equation \eqref{eq:Itotaleg3}, it is given by
\begin{equation}
\mathcal{T}_\mathrm{gl}=\frac{1}{6}\log \left( 1+\alpha^2+2\alpha\gamma\rho+\gamma^2 \right)\,.
\end{equation}
This quantity is actually increasing with correlation $\rho$ between sources, reflecting explicitly here that this is not a measure of complexity that operationalises inhomogeneity of information sources. That the information flow into the target is greatest when sources are strongly positively correlated is explained as follows: fluctuations of the sources cause fluctuations of the target, and fluctuations coming from positively correlated sources will more often combine to cause greater fluctuations of the target than of sources, whereas fluctuations coming from uncorrelated sources will more often cancel out at the target. Thus the relative variance of the target before compared with after knowing the pasts of the sources is greatest when sources are strongly positively correlated.

Conceptualising complexity as having to do with a whole system being greater than the sum of its parts, average synergistic information contributed by the past of a pair of variables to the present of a third variable could form a measure of complexity, by measuring the extent to which joint information contributed by two sources exceeds the sum of informations contributed by individual sources. Thus we could define the synergistic complexity $\mathcal{SC}$ as
\begin{equation}
\mathcal{SC}(\bM)=:\frac{2}{n(n-1)(n-2)} \sum_{i,j,k}  S(M_{i,t};\bM_{j,t}^-,\bM_{k,t}^-)\,.
\end{equation}
For Example 3, this leads via equation \eqref{eq:SEg3} to
\begin{equation}
\mathcal{SC}(\bM)=:\frac{1}{6} \log \left( 1+\alpha^2[1-\rho^2] \right)\,,
\end{equation}
for the case $|\alpha|\leq|\gamma|$, reflecting the strength and level of independence of the weakest connection. This is in the spirit of what the `$\Phi$' measures of integrated information \cite{Balduzzi2008,Barrett2011,Seth:2011} are supposed to capture (in some cases of high synergy `$\Phi$' measures are unsuccessful at doing this \cite{Griffith2014}). One could also conceive an analogous measure based on net synergy, but this does not lead to a formula that summarizes the complexity of Example 3 in any straightforward conceptualisation (see equation \eqref{eq:WMSEg3} for the non-zero term).

To fully understand the pros and cons of these various measures of complexity, they should be considered on systems composed of many (i.e.~$>>3$) elements. While there have been studies of causal density \cite{Seth:2011} and global transfer entropy \cite{Barnett13}, the synergistic complexity is a new measure, which will be explored in a follow up study, in controlled comparison with the other measures. One could further imagine, for general systems of $n$ variables, a complexity measure based on the synergistic information contributed to one variable from the pasts of all $(n-1)$ other variables. We do not attempt to consider such a measure here, since consideration of PIDs for more than two source variables is beyond the scope of this paper. This will also be an avenue for future research.
\\

\section{Discussion}

\subsection{Summary}

In this paper we have carried out analyses of partial information decompositions (PIDs) for Gaussian variables. That is, we have explored how the information that two source variables carry about a target variable decomposes into unique, redundant and synergistic information. Previous studies of PIDs have focused on systems of discrete variables, and this is the first study that focuses on continuous random variables. We have demonstrated that net synergy (i.e. the combined information being greater than the sum of the individual informations) is prevalent in systems of Gaussian variables with linear interactions, and hence that PIDs are non-trivial for these systems. We illustrated two interesting examples of a jointly Gaussian system exhibiting net synergy: (i) a case in which the target is correlated with both sources, but the two sources are uncorrelated (Fig.~\ref{fig:static_eg}(a)); (ii) a case in which the target is only correlated with one of two sources, but the two sources are correlated (Fig.~\ref{fig:static_eg}(b)). Further we have shown that, depending on the signs of the correlations between sources and target, net synergy can either increase or decrease with (absolute) correlation strength between sources (Fig.~\ref{fig:wms1}). Thus, redundancy should not be considered a reflection of correlation between sources.

Our key result is that for a broad class of Gaussian systems, a broad class of PIDs lead to: (i) a definition of redundancy as the minimum of the mutual informations between the target and each individual source, and hence they take redundancy as totally independent of the correlation between sources; (ii) synergy being the extra information contributed by the weaker source when the stronger source is known. Specifically, this holds for a jointly Gaussian system with a univariate target and sources of arbitrary dimension, and any PID for which the redundant and unique information depend only on the pair of marginal distributions of target and source 1 and target and source 2. This property has been argued for in \cite{Bertschinger13} and covers three previously proposed PIDs \cite{Williams10,Harder12,Griffith12,Bertschinger13}, which all operationalise distinct conceptualisations of redundancy (see Section \ref{sec:prevPIDs}). Thus it would be reasonable to apply this formula for redundancy to any data that are approximately Gaussian. Note however, there is still debate about the list of axioms a PID should satisfy beyond the minimal ones described in the Introduction \cite{Griffith13}, so it is still possible that an alternative PID is constructed for which the formula doesn't hold. We have termed the obtained decomposition the `Minimum Mutual Information' (MMI) PID. Most usefully, it is applicable in a multivariate time-series analysis to the computation of synergistic and redundant information arising in an arbitrary length past history of two variables about the present state of a third variable, i.e.~to analyses of information transfer.

That there can be net synergy when sources are uncorrelated implies that simple dynamical Gaussian systems can exhibit net synergy when considering the past of two variables as the sources and the present of one variable as the target. Indeed we have demonstrated this explicitly via some simple examples. We analyzed an MVAR model on which the pasts of two sources influence the present of a target (Fig.~\ref{fig:dynamic_eg}(c)), and showed that the synergistic information of the past of the sources about the target, as obtained via the MMI PID, increases monotonically with the weaker connection strength, and decreases monotonically with correlation between sources \eqref{eq:SEg3}. Thus, while redundancy doesn't provide us with distinct knowledge of the system, above and beyond mutual information between individual sources and target, synergy provides an intuitive formula for the extent of simultaneous differentiation (between sources) and integration (of information from both sources).

\subsection{Application to neuroscience}

Information theoretic analyses are increasingly popular in neuroscience, notably for analyzing the neural encoding of stimuli, or for analysing brain connectivity via quantification of information transfer between pairs of brain variables (especially if one considers Granger causality \cite{Wiener56,Granger69} as a measure of information transfer based on its correspondence with transfer entropy \cite{Barnett:2009a,Bressler11,Friston13}), see \cite{Wibral15} for a recent review. There have been several studies in which net synergy/redundancy has been computed empirically on neurophysiological datasets, e.g.~\cite{Brenner00,Machens01,Bettencourt07,Bettencourt08,Stramaglia12,Gaucher13,Marinazzo14}. In neural coding, net synergy (WMS$>0$) has been observed in the information successive action potentials carry about a stimulus \cite{Brenner00}. In most studies, information transfer between EEG variables has tended to exhibit net redundancy (i.e.~WMS$<0$), although recently net synergy (WMS$>0$) has been observed in information transfer amongst some intracranial EEG variables in an epileptic patient \cite{Stramaglia14}. A pair of recent studies has associated certain pathological brain states with increased net redundancy in information transfer: amongst electrocorticographic time-series (contacts placed intracranially on the surface of the cortex) during seizure onset in an epileptic patient \cite{Stramaglia12}; and amongst scalp EEG time-series from traumatic brain injury patients in the vegetative state, compared to analogous recordings from healthy controls \cite{Marinazzo14}.

Usually net redundancy has been assumed to arise due to common sources, and hence correlation between variables. However, as mentioned above, the results here suggest that this is not always the case. For the Gaussian case we have considered, this holds for positive correlation between sources and an equal correlation between the target and each of the sources, but not more generally (see Fig.~\ref{fig:wms1}).

The canonical example scenario for net synergy takes one of the sources to be a ``suppressor" variable, entering a regression via a multiplicative term with the other source \cite{Stramaglia14}. Such non-linear systems are non-Gaussian, so PID on systems with suppressor variables is beyond the scope of this paper. However, our demonstration of cases of net synergy for linear Gaussian systems suggests that observing net synergy does not necessarily imply the presence of a suppressor variable. Further, in concordance with the non-straightforward relationship found here between net synergy and correlation between sources, it has been shown in \cite{Nirenberg03} and \cite{Latham05} that net synergy is not a useful measure for assessing the importance of correlations between neurons (or neural populations) for successful stimulus decoding.

Using the MMI PID, redundancy and synergy can now be computed separately on neurophysiological datasets on which a Gaussian approximation is valid to bring more detailed insight into information theoretic analyses.

%

\subsection{Final remarks}

We found that if one were to quantify information as reduction in variance rather than reduction in entropy for jointly univariate Gaussian variables, then the net synergy would be precisely zero for uncorrelated sources (see Section \ref{sec:prev}). Since it is counterintuitive that synergy should arise in the absence of interactions between sources, this suggests that perhaps reduction in variance is a better measure of information for Gaussian variables than mutual information based on Shannon entropy, which results in information being based on the concave log function, and leads to a distorting effect when comparing combined information from two sources with the sum of information from each source on its own in the formula for net synergy. Since Shannon information between continuous random variables is more precisely based on differential, as opposed to absolute entropy (see Section \ref{sec:prelims}), its interpretation in terms of reduction of uncertainty is in any case somewhat ambiguous, in spite of being widely used. One would however lose the symmetry of information if redefining it as reduction in variance. Angelini et al.~\cite{Angelini10} made a similar observation for Granger causality: a formula based solely on variances, without taking logarithms, results in the Granger causality from a group of independent variables being equal to the sum of Granger causalities from the individual variables (assuming linearity). Future studies of synergy might benefit from further consideration of alternative measures of basic mutual information for continuous random variables.

The MMI PID constitutes a viable candidate PID for information sharing and transfer amongst a group of three jointly Gaussian variables. This will be useful given that the Gaussian approximation is so widely used when analysing continuous time-series. There is therefore the possibility of application of the MMI PID to a broad range of complex systems, opening up the opportunity to explore relations between any macroscopic phenomenon and the distinct categories of information sharing (redundant, unique and synergistic) amongst triplets of continuous time-series variables. The isolation of synergistic information from the other categories could be useful for measuring complexity, by quantifying more correctly than difference in mutual information alone, the extent to which information from multiple sources taken together is greater than that from individual sources taken separately (see Section \ref{sec:complexity}). A challenge for future work is to obtain a more general framework for PIDs on continuous random variables: for variables following other distributions, and for the scenario of more than two source variables.

\section*{Acknowledgements}
I thank Lionel Barnett, Joseph Lizier and Michael Wibral for invaluable discussions during the writing of this paper, Anil Seth for very useful comments on draft manuscripts, Michael Schartner for a read-through of the final draft before submission, and Johannes Rauh for comments on the first ArXiv version. I thank Marcello Massimini and his laboratory for hosting me as a visiting fellow at the University of Milan. ABB is funded by EPSRC grant EP/L005131/1.

\end{document}